\documentclass[%
reprint,
superscriptaddress,
showpacs,
 amsmath,amssymb,
 aps,
prb,
]{revtex4-1}

\usepackage{graphicx}
\usepackage{dcolumn}
\usepackage{bm}

\begin{document}

\preprint{APS/123-QED}

\title{Magnetic vortex lattice in HgBa$_2$CuO$_{4+\delta}$ observed by small-angle neutron scattering}

\author{Yuan~Li}
\email[Electronic address: ]{yuan.li@fkf.mpg.de}
\affiliation{Department of Physics, Stanford University, Stanford, California 94305, USA}
\affiliation{Max Planck Institute for Solid State Research, D-70569 Stuttgart, Germany}
\author{N.~Egetenmeyer}
\author{J.~L.~Gavilano}
\affiliation{Laboratory for Neutron Scattering, ETH Zurich and Paul Scherrer Institute, CH-5232 Villigen, Switzerland}
\author{N.~Bari\v{s}i\'{c}}
\altaffiliation[Present address: ]{School of Physics and Astronomy, University of Minnesota, Minneapolis, Minnesota 55455, USA}
\affiliation{1. Physikalisches Institut, Universit\"{a}t Stuttgart, D-70550 Stuttgart, Germany}
\author{M. Greven}
\affiliation{School of Physics and Astronomy, University of Minnesota, Minneapolis, Minnesota 55455, USA}

\date{\today}

\begin{abstract}
We report a direct observation of the magnetic vortex lattice in the model high-temperature superconductor HgBa$_2$CuO$_{4+\delta}$. Using small-angle neutron scattering on high-quality crystals, we observe two equal domains of undistorted triangular vortex lattices well-aligned with the tetragonal crystallographic axes. The signal decreases rapidly with increasing magnetic field and vanishes above $0.4\,$Tesla, which we attribute to a crossover from a three-dimensional to a two-dimensional vortex system, similar to previous results for the more anisotropic compound Bi$_{2.15}$Sr$_{1.95}$CaCu$_2$O$_{8+\delta}$. Our result indicates that a triangular vortex lattice at low magnetic fields is a generic property to cuprates with critical temperatures above $80\,$K.
\end{abstract}

\pacs{74.25.Uv, 61.05.fg, 74.72.Gh}
\maketitle


\section{Introduction}
The cuprate high-temperature superconductors exhibit a rich variety of mesoscopic phenomena associated with magnetic vortices in the mixed state.\cite{BlatterPRL94} In clean samples, the vortices lower their energy by forming a long-range-ordered vortex lattice (VL), which provides an opportunity to study the superconductivity of the bulk. In a simplified picture,\cite{Abrikosov57} the vortex physics is determined by two parameters: the penetration depth $\lambda$ of magnetic field into the superconducting phase and the coherence length $\xi$ of the superconducting order parameter. Since both of these parameters exhibit pronounced anisotropies in the cuprates, the VL is expected to reflect important characteristics of the superconductivity, such as the $d$-wave pairing symmetry.\cite{AffleckPRB97,IchiokaJPSJ97,IchiokaPRB99,ShiraishiPRB99} However, there exists a substantial variety of VL properties among different compounds, including Bi$_{2.15}$Sr$_{1.95}$CaCu$_2$O$_{8+\delta}$ (Bi2212, Ref.~\onlinecite{CubittNature93}), YBa$_2$Cu$_3$O$_{6+\delta}$ (YBCO, Refs.~\onlinecite{BrownPRL04,WhitePRL09}), and La$_{2-x}$Sr$_x$CuO$_4$ (LSCO, Ref.~\onlinecite{GilardiPRL02}). The situation is further complicated by the presence of twin domains and boundaries in the most studied compound YBCO.\cite{YethirajPRL93,KeimerScience93,JohnsonPRL99} Therefore, it is highly desirable to measure a structurally-simple representative system in order to distinguish the generic vortex physics from material-specific aspects.

We report here the first observation of the VL in HgBa$_2$CuO$_{4+\delta}$ (Hg1201) using small-angle neutron scattering (SANS).  Hg1201 is in many ways a model system.\cite{EisakiPRB04,BarisicPRB08} First, Hg1201 exhibits the highest superconducting critical temperature ($T_\mathrm{c}$) among all cuprates with a single CuO$_2$ layer per primitive unit cell. Notably, our samples exhibit the highest $T_\mathrm{c}$ among all materials in which VL studies by SANS have been carried out to date. Second, the CuO$_2$ layer of Hg1201 is free of long-range structural distortions,\cite{BalagurovPRB99} and disorder resides in the Hg-O layer, relatively far away from the CuO$_2$ layer.\cite{WagnerPhysicaC93,PelloquinPhysicaC97,VialletPhysicaC97}  As disorder and defects are common causes for vortex pinning, their relative absence in Hg1201 results in weak pinning\cite{BarisicPRB08} and is important for the formation of the VL. Third, Hg1201 bridges some differences between other cuprates: LSCO possesses a single-layer structure and relatively low $T_\mathrm{c}$, whereas YBCO and Bi2212 are double-layer compounds with a high $T_\mathrm{c}$. On the other hand, Hg1201 possesses a high $T_\mathrm{c}$ and it is a single-layer compound. Finally, while the other three compounds exhibit long-range structural distortions in the superconducting doping range, the high tetragonal symmetry of Hg1201 (and the absence of long-range distortion) allows us to avoid complications arising from vortex-pinning due to crystallographic twin boundaries.

\section{Experimental methods}
High-quality Hg1201 crystals have recently become available and proven to be suitable for cutting-edge neutron scattering measurements.\cite{LiNature08,YuPRB10,LiNature10} The current study was carried out on two arrays (total mass of approximately 60 and $90\,$mg) of nearly optimally-doped Hg1201 crystals. The crystals were grown by a self-flux method\cite{ZhaoAdvMater06} and subsequently heat-treated\cite{BarisicPRB08} for up to 180 days in order to reach the desired homogeneous hole concentration. They were further selected according to the results of magnetometery measurements to ensure high quality,\cite{BarisicPRB08} and had a uniform $T_\mathrm{c}$ of 94$\,$K (determined from the transition mid-point) with a width of less than 2$\,$K. The crystals were co-aligned and glued on a silicon wafer using an x-ray Laue diffractometer. The $c$-axes of the crystals were within $1^\circ$ perpendicular to the surface of the wafer, and the $a$- and $b$-axes were co-aligned to within $1.5^\circ$. Since the two samples gave essentially the same result, we do not distinguish between them in the remainder of this paper. The measurements were performed on the SANS-I spectrometer at the Paul Scherrer Institute, Switzerland. Incident neutrons of wavelengths ($\lambda_\mathrm{n}$) of 5--10$\,\mathrm{\AA}$ were selected and collimated over a distance of 8--18$\,$m before the sample. Diffracted neutrons were observed by a two-dimensional multidetector located 13--20$\,$m after the sample. The sample was mounted in a horizontal cryomagnet with the $a$-axis vertical and the $c$-axis pointing along the direction of the applied magnetic field (within 1$^\circ$). The VL was prepared by cooling the sample from above $T_\mathrm{c}$ in a field oscillating around the target value (by $\sim1\%$ of the target value). This method\cite{WhitePRL09} improved the VL ordering by keeping the vortices away from local pinning potentials. After reaching the lowest temperature of 2$\,$K, the field was held stationary during diffraction measurements, which involved rotating the sample and cryomagnet together to angles that brought various diffraction peaks onto the Bragg condition at the detector.  Measurements at temperatures between 2$\,$K and $T_\mathrm{c}$ were performed upon heating without changing the field.  For all the data presented here, the background measured at $T=2\,$K in zero field or at $T>T_\mathrm{c}$ has been subtracted from the field-cooled measurements to extract the VL signal.

\section{Results}
\begin{figure}
\includegraphics[width=3.375in]{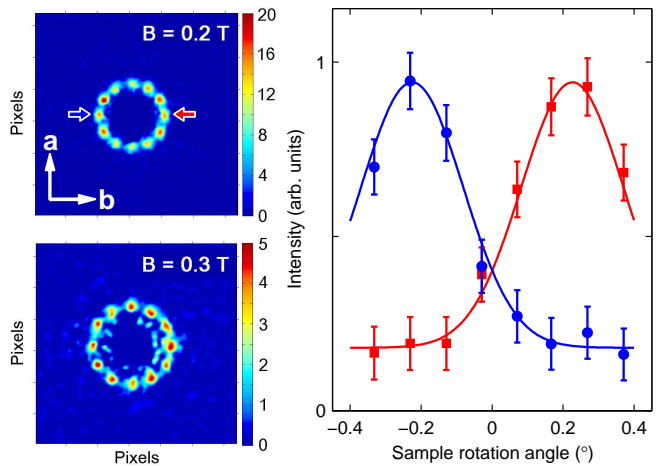}
\caption{\label{fig:one}
Left: Neutron diffraction patterns at $B=0.2$ and $0.3\,$T and $T=2\,$K, averaged over different sample rotation angles. Right: Rocking scans for two opposing Bragg reflections (indicated by the arrows on the left), measured by rotating the sample-magnet assembly with respect to the incident neutron beam.
}
\end{figure}

Figure~\ref{fig:one} shows diffraction patterns collected at $T=2\,$K for $B=0.2$ and $0.3\,$T. The data are averaged over multiple sample rotation angles to display all first-order diffraction peaks. Given the tetragonal crystallographic symmetry of Hg1201, the VL is expected to form two equally-populated domains aligned with the $a$- and $b$-axes. The presence of twelve ($2\times6$) diffraction peaks with comparable intensities confirms this, and indicates that the VL is triangular. Well-defined peaks are observed in rocking scans by rotating the sample-magnet assembly, which suggests that the magnetic flux lines are straight and parallel to the $c$-axis.

\begin{figure}
\includegraphics[width=3.375in]{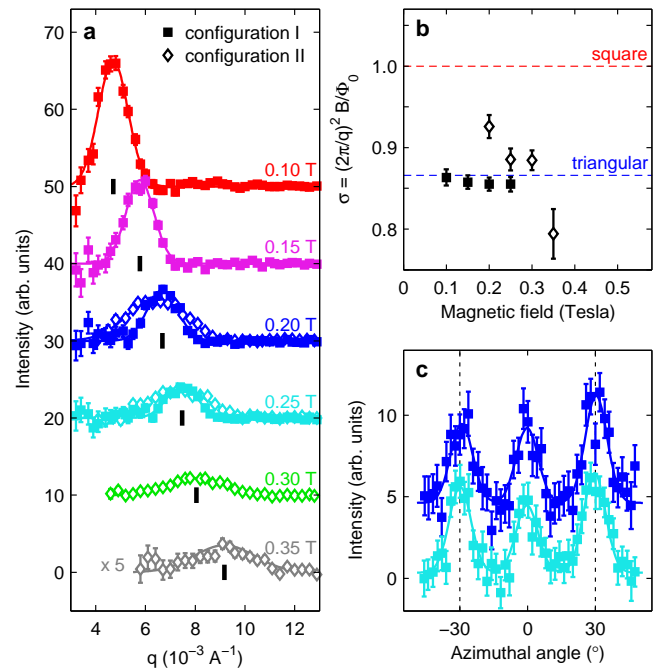}
\caption{\label{fig:two}
(a) Azimuthally-averaged signal amplitudes as a function of momentum transfer $q$, measured at $T=2\,$K in different magnetic fields. Data obtained with two configurations (see text) are rescaled to allow a common vertical axis. The $B=0.35\,$T data are furthermore multiplied by a factor of 5 for clarity. Vertical bars indicate the fitted value of the characteristic $q$. (b) The VL-structure-dependent coefficient $\sigma$ (see text) at different fields. Filled and open symbols correspond to configurations I and II, respectively. (c) Azimuthal-angle dependence of the signal measured in configuration I at $T=2\,$K for $B=0.2$ and 0.25$\,$T. Data are vertically offset for clarity.
}
\end{figure}

The radial intensity distribution of the VL signal is displayed in Fig.~\ref{fig:two}a. Two spectrometer configurations were used in order to maximize the signal-to-background ratio at different fields: collimation of 18$\,$m with $\lambda_\mathrm{n} = 10\,\mathrm{\AA}$ (configuration I) and collimation of 11$\,$m with $\lambda_\mathrm{n} = 8\,\mathrm{\AA}$ (configuration II), where the sample-detector distance was equal to $20\,$m in both cases. The characteristic wave vector magnitude $q$ of the VL is extracted by fitting the data to a Gaussian peak. This quantity is related to the magnetic field through the relation $\sigma=(2\pi/q)^2B/\Phi_0$, where $\Phi_0$ is the flux quantum $h/2e$, and $\sigma$ is a geometric factor equal to $\sqrt{3}/2$ or 1 for triangular or square VL, respectively. As can be seen from Fig.~\ref{fig:two}b, the measured values of $\sigma$ are consistent with a triangular VL up to $B=0.35\,$T (even though the determination of $q$ using configuration II has somewhat large errors because of the relatively poor momentum resolution). The geometry of the VL can also be determined from the azimuthal intensity distribution near the characteristic $q$. For $B=0.2$ and 0.25$\,$T, the nearest diffraction peaks from the same type of domain are precisely 60$^\circ$ apart (Fig.~\ref{fig:two}c), which is evidence that the triangular VL is undistorted. The fact that no intensity is observed between the peaks indicates that the VL domains are nearly fully aligned with the crystallographic axes.


\begin{figure}
\includegraphics[width=3.375in]{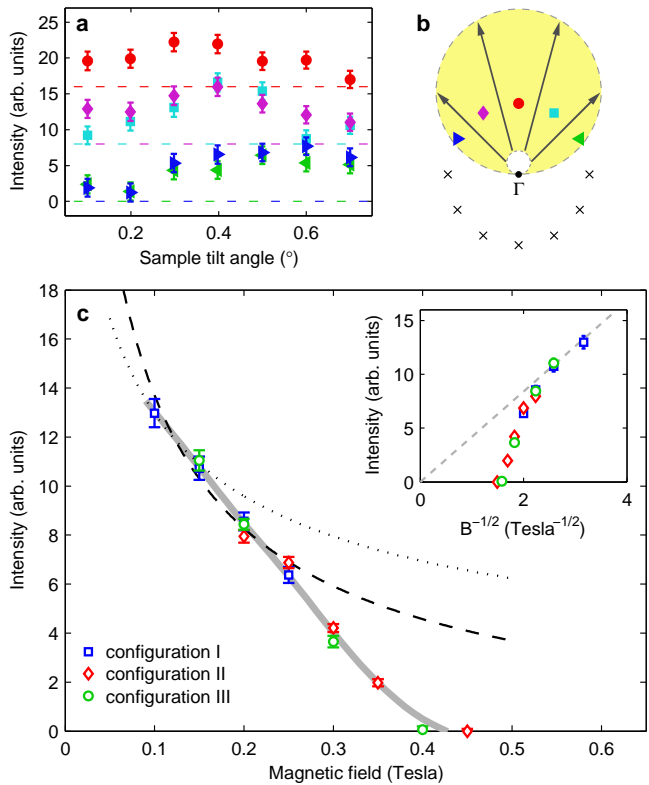}
\caption{\label{fig:three}
(a) Intensities measured at $T=2\,$K and $B=0.25\,$T in sector boxes surrounding the diffraction peaks indicated by corresponding symbols in (b). The data are offset for clarity, and the same offset is used for reflections that are symmetric about the vertical plane. (b) Illustration of a tilt scan (see text). Broken circles are the intersections of the Ewald sphere with the two-dimensional reciprocal plane at the end points of the scan. Potential diffraction spots are marked by crosses. The five diffraction spots marked by symbols have their Bragg conditions satisfied within the range of the scan. (c) VL intensity measured at $T=2\,$K as a function of magnetic field. Measurements in different configurations are normalized to the values at common magnetic fields. The dotted line is the expected field dependence in the London limit (Eq.~\ref{formfactor}); the dashed line is the expected behavior after vortex-core correction (see text), using $B_\mathrm{c2}=100\,$T (Ref.~\onlinecite{LeBrasPhysicaC96}) and the phenomenological parameter $s=7$; the solid line is a smoothing curve describing the data. The inset shows the same data plotted versus $B^{-1/2}$, where the field dependence expected in the London limit is represented by the dashed line.
}
\end{figure}

While it can be seen from Fig.~\ref{fig:two}a that the signal decreases rapidly with increasing magnetic field, the azimuthally-averaged intensity does not provide the most accurate determination of the signal amplitude, because the Bragg conditions for different diffraction spots are not simultaneously satisfied. In order to determine the field dependence of the signal, we employed a procedure that is illustrated in Fig.~\ref{fig:three}b: for each of the several magnetic fields measured in a given configuration, a scan is performed on the tilt angle of the sample-magnet assembly, which can be viewed as rotating the Ewald sphere about the $\Gamma$ point with respect to the two-dimensional reciprocal plane of the VL. By definition, both the Ewald sphere and the two-dimensional reciprocal plane contain the $\Gamma$ point, and their intersection is a circle which expands as the tilt angle increases (arrows in Fig.~\ref{fig:three}b). When this circle goes through a diffraction spot, the intensity of that spot is maximized, and the different spots have their intensities maximized at different tilt angles, as can be seen from Fig.~\ref{fig:three}a. To measure the signal amplitude of the VL, we take the average of the maximal intensities of the five spots on the half of the reciprocal plain that corresponds to the tilting direction (for the $B=0.25\,$T example, the averaged value of the five maxima in Fig.~\ref{fig:three}a). In addition to configurations I and II, a third configuration III (collimation of 18$\,$m with $\lambda_\mathrm{n} = 8\,\mathrm{\AA}$ and sample-detector distance $=18\,$m) was used as a consistency check, and very good agreement is found among the data sets obtained in different configurations. The result is summarized in Fig.~\ref{fig:three}c.

The intensity $I_{hk}$ of VL reflection $(h,k)$ is given by\cite{ChristenPRB77}
\begin{equation}\label{intensity}
I_{hk}=2\pi \phi \left(\frac{\mu}{4}\right)^2 \frac{V \lambda_\mathrm{n}^2}{\Phi_0^2 q_{hk}} \left| F_{hk}\right|^2 \propto \frac{\left| F_{hk}\right|^2}{q_{hk}},
\end{equation}
where $\phi$ is the incident neutron flux, $\mu$ is the neutron magnetic moment, $V$ is the sample volume, and $q_{hk}$ is the magnitude of the wave vector of the reflection. $F_{hk}$ is the form factor, which is a Fourier component of the spatial variation of the magnetic field. In the London limit (\textit{i.e.}, without considering the effect of vortex cores), $F_{hk}$ is related to the penetration depth $\lambda$ by
\begin{equation}\label{formfactor}
F_{hk}=\frac{B}{1+\left(q_{hk} \lambda \right)^2}.
\end{equation}
For $B \gg B_\mathrm{c1}$ (the lower critical field), the second term in the denominator is dominant and this gives $I_{hk}\propto 1/q_{hk} \propto B^{-1/2}$ (since for a given VL symmetry, $q_{hk}\propto \sqrt{B}$). In type-II superconductors, the presence of the normal-state vortex cores further reduces the VL signal at high fields,\cite{YaouancPRB97} an effect that can be approximated by multiplying Eq.(\ref{formfactor}) with the correction factor $\displaystyle e^{-s\cdot B/B_\mathrm{c2}}$, where $s$ is a phenomenological parameter on the order of unity and $B_\mathrm{c2}$ is the upper critical field. As can be seen from Fig.~\ref{fig:three}c, the measured intensity decreases with increasing field more rapidly than $\propto B^{-1/2}$ above $B\approx0.15\,$T (inset) and more rapidly than the estimated behavior after vortex-core correction above $0.25\,$T. The signal becomes undiscernibly small above 0.35$\,$T, and measurements at higher fields (up to 4.0$\,$T) do not indicate a reappearance of the signal (not shown).

The disappearance of the VL signal near $B=0.4\,$T can be explained by a mechanism\cite{CubittNature93} first proposed for Bi2212: in an anisotropic superconductor, magnetic vortices are pinned over a characteristic length along the $c$-axis that is related to both the field strength and the anisotropy of the penetration depth, $\gamma\equiv(\lambda_{\mathrm{c}} / \lambda_\mathrm{ab})$. At sufficiently high fields, this length decreases below the distance between the adjacent CuO$_2$ layers, and the VL decomposes into ``vortex pancakes'' that are two-dimensionally ordered in each CuO$_2$ layers but are uncorrelated along the $c$-axis.  This substantially reduces the spatial variation of the magnetic field and suppresses the diffraction signal. Such a dimensional crossover is expected to occur at $B_\mathrm{2D} \sim \Phi_0/(\gamma c)^2 \sim 1.4\,$T for Hg1201, where $c=9.6\,\mathrm{\AA}$ is the $c$-axis lattice constant and $\gamma$ is $\sim 40$ (Ref.~\onlinecite{PanagopoulosPRB03}). Both the estimated value of $B_\mathrm{2D}$ and the measured value of $0.4\,$T are substantially lower than $B_\mathrm{c2}=100\,$T,\cite{LeBrasPhysicaC96} and than the magnetic field ($B=10.8\,$T) up to which the VL has been observed in the more isotropic compound YBCO,\cite{WhitePRL09} where $B_\mathrm{2D}$ is expected to be much higher than in Hg1201. The difference between the estimated and measured values of $B_\mathrm{2D}$ might arise because $\gamma$ only serves as a phenomenological parameter in determining $B_\mathrm{2D}$, and because there is considerable uncertainty in the measured value of $\gamma$ associated with doping\cite{PanagopoulosPRB03} and the experimental method.\cite{PanagopoulosPRL97,PanagopoulosPRB03,LeBrasPhysicaC96,GuSupercondSciTechnol98} We note that the width of the rocking curves in our measurement (Fig.~\ref{fig:one}) suggests that the magnetic flux lines are straight over long distances. This finding is similar to the case of Bi2212 (Ref.~\onlinecite{CubittNature93}) and can be reconciled with the dimensional-crossover mechanism for the VL decomposition if the flux lines are distorted only over short distances. In other words, as the flux lines break down into vortex pancakes, the magnetic field profile projected onto the basel plane dramatically loses its contrast, but the correlation between the averaged location of the flux lines (or stacks of vortex pancakes) remains intact. Finally, for the vortex pancakes that belong to a single flux line to be misaligned along the $c$-axis, the pancakes in different CuO$_2$ layers do not need to be individually pinned; instead, neighboring pancakes in the same CuO$_2$ layer may be effectively pinned together if the two-dimensional correlation length of the VL is larger than the size of each pancake and than the characteristic distance between neighboring pinning centers in the same layer. Given the well-defined diffraction patterns in Figs.~\ref{fig:one} and \ref{fig:two}c, it is likely that the vortex pancakes are pinned by a low density of pinning centers in this way.

\begin{figure}
\includegraphics[width=3.3in]{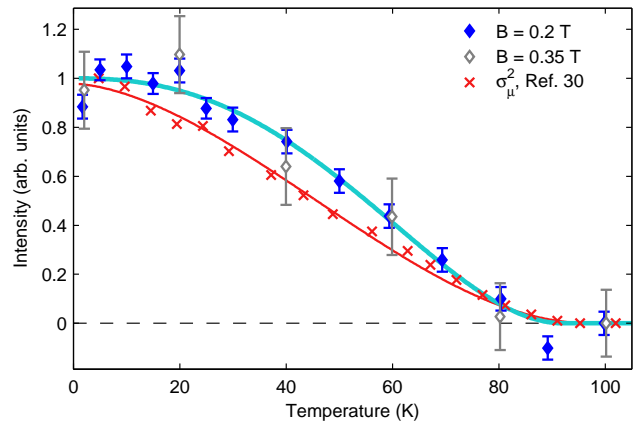}
\caption{\label{fig:four}
Temperature dependence of diffraction intensity measured at two different magnetic fields compared with $\mu$SR results for optimally-doped Hg1201.\cite{NachumiHyperfine97} The value at $T=100\,$K is set to zero, and the data are normalized to the values at low temperature. The solid lines are empirical fits as guides to the eye (see text).
}
\end{figure}

The temperature dependence of the VL signal was measured in two magnetic fields $B=0.2$ and $0.35\,$T. For each field, the measurement was performed at a fixed sample angle, and the intensities of multiple diffraction peaks were summed together in order to improve the counting statistics. Apart from a rescaling factor for the overall intensity, the temperature dependence measured in the two fields is identical (Fig.~\ref{fig:four}). Since $B=0.35\,$T is very close to the field at which the VL signal vanishes, this suggests that the decrease of signal as $T_\mathrm{c}$ is approached (due to the increasing penetration depth, $I\propto \lambda^{-4}$) and the disappearance of the signal at $T=2\,$K above $B=0.4\,$T have different causes. The similar temperature dependence in the two fields further supports the dimensional-crossover scenario of the VL decomposition, since $B_\mathrm{2D}$ is related to the anisotropy $\gamma$ which is nearly independent of temperature.\cite{LeBrasPhysicaC96}

The data in Fig.~\ref{fig:four} can be described by the empirical function $I \propto \left[1-\left(T/T_\mathrm{c}\right)^\alpha\right]^2$. The fitted value for $T_\mathrm{c}$ is $92\pm5\,$K, consistent with the $T_\mathrm{c}$ value of our sample determined by magnetic susceptibility measurements ($94 \pm 1\,$K). The fitted value for $\alpha=2.4\pm0.3$ is rather different from that of the ``two-fluid'' empirical expression ($\alpha=4$) for conventional superconductors, which seems to describe early $\mu$SR results for YBCO.\cite{HarshmanPRB89} (In $\mu$SR experiments, the measured muon relaxation rate $\sigma_\mu$ is proportional to $\lambda^{-2}$, and therefore $\sigma_\mu^2$ is expected to be proportional to the signal amplitude measured by SANS.) The data are also somewhat different from the more recent $\mu$SR result for Hg1201,\cite{NachumiHyperfine97} where $\sigma_\mu^2$ can be described by the same function with $\alpha=1.6$ (Fig.~\ref{fig:four}). These differences need to be further understood. We note that a discrepancy between SANS and $\mu$SR results regarding the temperature dependence has been previously documented,\cite{ForganNature90} and the temperature dependence in our data is consistent with other SANS measurements in the cuprates.\cite{CubittNature93,ForganNature90}  For $d$-wave superconductors like the cuprates, a more quantitative description of the data near $T=0\,$K may require a correction to account for nonlocal effects.\cite{KosztinPRL97}

\section{Discussion and conclusion}

Now we discuss the implication of our results for the VL physics in the cuprates. So far, well-defined VLs at $B>1\,$T have been observed only in YBCO and LSCO,\cite{BrownPRL04,WhitePRL09,GilardiPRL02} and the results are drastically different: in YBCO,\cite{WhitePRL09} a hexagonal VL pattern is observed at low fields, and a square pattern is observed only at $B > 6.7\,$T, with the nearest-neighbor direction along the Cu-Cu diagonal of the unit cell. In contrast, a square pattern is already well developed at $B=0.4\,$T in LSCO,\cite{GilardiPRL02} and the nearest-neighbor direction is along the Cu-O direction, $45^\circ$ from that in YBCO. A square VL has also been observed in the electron-doped cuprate Nd$_{2-x}$Ce$_x$CuO$_4$ (NCCO, Ref.~\onlinecite{GilardiPRL04}) down to $B=0.05\,$T. Our result for Hg1201 indicates that a unified picture may still exist for cuprates with high values of $T_\mathrm{c}$ ($>80\,$K), in which a triangular VL is favored at low fields ($B<0.5\,$T). We note that both Hg1201 and YBCO have non-body-centered crystal structures. Given that the formation of a square VL in YBCO is likely driven by the anisotropy of the superconducting order parameter,\cite{WhitePRL09} the square VLs at low fields in LSCO and NCCO do not seem to have the same origin, but may instead be related to their relatively low $T_\mathrm{c}$ values and/or to their common body-centered crystal structure. To further test the validity of this picture, measurements of the VL in Hg1201 and Bi2212 at higher fields (such as using local probes) are desirable.

In conclusion, we have used SANS to observe the VL in Hg1201 near optimal doping. Our data suggest that a triangular VL at low fields ($B<0.5\,$T) is generic to cuprates with high $T_\mathrm{c}$ and non-body-centered structure. The difficulty in using SANS to measure VLs at high fields in this anisotropic superconductor is likely due to a decomposition of magnetic flux lines into two-dimensional vortex pancakes.

\begin{acknowledgments}
We thank Dmytro S. Inosov and Guichuan Yu for stimulating discussions. This work was supported by the DOE under Contract No. DEAC02-76SF00515 and by the NSF under Grant No. DMR-0705086.  Y.L. and N.B. acknowledge support from the Alexander von Humboldt foundation.
\end{acknowledgments}

\bibliography{sans_final}

\end{document}